# Low-Voltage Magnetoelectric Coupling in Membrane Heterostructures

S. Lindemann,[1,†] J. Irwin,[2,†] G.-Y. Kim,[3] B. Wang,[4] K. Eom,[1] J. J. Wang,[4] J. M. Hu,[1] L. Q. Chen,[4] S. Y. Choi,[3] C. B. Eom,[1*] M. S. Rzchowski,[2,*]

[1] Department of Materials Science and Engineering, University of Wisconsin-Madison Madison, Wisconsin 53706, United States.

[2] Department of Physics, University of Wisconsin-Madison Madison, Wisconsin 53706, United States.

[3] Department of Materials Science and Engineering, Pohang University of Science and Technology, Pohang, Gyeongbuk 37673, Korea.

[4] Department of Materials Science and Engineering, Pennsylvania State University, University Park, Pennsylvania 16802, United States.

**Strain-mediated magnetoelectric (ME) coupling in ferroelectric (FE) / ferromagnetic (FM) heterostructures offers a unique opportunity for both fundamental scientific research and low power multifunctional devices. Relaxor-ferroelectrics, like $(1-x)Pb(Mg_{1/3}Nb_{2/3})O_3$-$(x)PbTiO_3$ (PMN-xPT), are ideal FE layer candidates due to their giant piezoelectricity. But thin films of PMN-PT suffer from substrate clamping which substantially reduces piezoelectric in-plane strains. Here we present the first demonstration of low voltage ME coupling in an all-thin-film heterostructure which utilizes the anisotropic strains induced by the (011) orientation of PMN-PT. We completely remove PMN-PT films from their substrate and couple with FM Ni overlayers to create membrane PMN-PT/Ni heterostructures showing 90° Ni magnetization rotation with 3V PMN-PT bias, much less than the bulk PMN-PT ~100V requirement. Scanning transmission electron microscopy and phase-field simulations clarify the membrane response. These results provide a crucial step towards understanding the microstructural behavior of PMN-PT thin films for use in piezo-driven magnetoelectric heterostructures.**

[†]J.I. and S.L. contributed equally to this work.

[*]To whom correspondence should be addressed. E-mail: eom@engr.wisc.edu, rzchowski@physics.wisc.edu

**Introduction**

Electric-field control of magnetism, also known as converse magnetoelectric coupling, in ferromagnetic (FM) / ferroelectric (FE) composites is of significant interest due to the potential for its development as next generation memory storage and sensing technologies.(*1, 2*) Of particular interest for use as the FE layer are relaxor-ferroelectrics, such as (1-x)Pb(Mg$_{1/3}$Nb$_{2/3}$)O$_3$-(x)PbTiO$_3$ (PMN-PT), which show large piezoelectric response for compositions near a morphotropic phase boundary (MPB) (x = ~35% for PMN-PT).(*3*) By coupling the relaxor-ferroelectric with a FM containing large magnetostriction, converse magnetoelectric coupling is achieved through transfer of the voltage-induced strain from the FE layer into the FM layer which can result in strain-mediated control of in-plane magnetic anisotropy(*4–6*), tunneling magnetoresistance(*7*), ferromagnetic resonance(*8*), and conductivity(*9*).

For many strain-mediated magnetoelectric coupling applications, including rotation of the in-plane magnetization of a coupled FM, anisotropic in-plane strains are required. Therefore, the above-mentioned studies (*4-7,9*) used (011) oriented bulk single crystals of PMN-PT which develop large anisotropic in-plane strains under an electric field. The x = 30% composition of PMN-PT is rhombohedral (R) with spontaneous polarization along the <111> directions.(*10*) As shown in Fig. 1A, these polarization directions can be grouped as rhombohedral up (R$_{UP}$, blue), rhombohedral in-plane (R$_{IP}$, orange), and rhombohedral down (R$_{DOWN}$, not shown).(*4*) Additionally, applying a large electric field can stabilize a polarization parallel to the applied field direction [011], giving the crystal an Orthorhombic (O) up (O$_{UP}$, purple) symmetry.(*11*) Each of these polarization groups result in average strained unit cells projected into the (011) plane as shown in Fig. 1B, with the unstrained cubic cell (dashed lines) as a reference. This uses equal-weight averaging of each polarization vector present in the group. The normal strains associated with each polarization group may be calculated using the PMN-PT electrostriction tensor (see Supplementary Table S1), which has been measured in bulk.(*12*) Fig. 1C shows the normal strains $\varepsilon_{xx}$ and $\varepsilon_{yy}$ along the x and y directions, respectively, as well as the anisotropic in-plane strain $\varepsilon_{xx} - \varepsilon_{yy}$ for all three polarization groups with the same magnitude of polarization, again averaged over all ferroelectric domains in the polarization group. Studies using bulk PMN-PT primarily utilized 71°/109° permanent switching between R$_{UP}$ and the metastable R$_{IP}$ polarization states, though as seen in Fig. 1C, inducing polarization rotation from R to O can result in even larger strain anisotropy. The R to O transition is often deemed undesirable in bulk studies due to the large required voltage as well as it being a non-permanent effect, i.e., the strains will relax once the voltage bias is removed. In thin films, however, the high electric fields needed for polarization rotation between the R and O directions can be more easily achieved and plays a large role in this study.

The drive towards low-power magnetoelectric devices(*13, 14*), as well as development of micro- and nano-electromechanical systems, has prompted the study of relaxor-ferroelectric thin films.(*15–17*) Upon reduction to thin film dimensions, relaxor-ferroelectrics suffer a large reduction in piezoelectricity due to mechanical clamping by a passive substrate.(*18–23*) Such a limitation presents a significant challenge towards successful integration of relaxor-ferroelectric thin films in high-performance devices. Several methods have been used to reduce clamping in ferroelectric thin films including growth directly on flexible substrates(*24*), micromachining into bendable cantilevers(*25*), fine lithographic patterning(*26–28*), micromachining(*29*, *30*), and direct ME coupling for magnetic field sensing in thin film cantilevers.(*31*, *32*)  But obtaining the largest strain-mediated magnetoelectric response in thin-film heterostructures necessitates complete removal of the substrate to allow a free piezoresponse.

Many device concepts based on the piezo-driven ME effect rely on the use of (011)-oriented PMN-PT thin films due to the above-mentioned demonstrations involving bulk PMN-PT. Up to now, however, demonstration of piezo-driven ME coupling in all-thin-film FE/FM heterostructures has been hindered by the issue of substrate clamping. In this study, we overcome the clamping issue and provide first demonstration of low voltage strain-mediated ME coupling in an all-thin-film heterostructure that only relies on the anisotropic strains inherent to the (011) orientation of PMN-PT. We fabricate unclamped (011)-oriented PMN-PT thin film membranes, through release from a rigid substrate using a sacrificial etching layer, and couple them with FM Ni thin film overlayers. Using the symmetry-enabled piezostrains, we demonstrate robust 90° rotation of in-plane magnetic anisotropy in the Ni through application of just 3V applied bias across the PMN-PT, compared to the >100V required when using bulk single crystals of PMN-PT.(*4*) We find that the piezostrains exhibited by the PMN-PT membrane can be attributed to driving the PMN-PT film's polarization towards the O state under the applied bias, which similar to bulk behavior, results in a non-permanent effect. But contrary to bulk behavior, we do not see evidence of the film switching into a metastable $R_{IP}$ polarization state. To further understand the domain behavior of the PMN-PT membrane, we used high resolution Scanning Transmission Electron Microscopy (STEM) to map b-site cation displacements in order to observe the domain configuration of the PMN-PT membrane. The STEM imaging shows the membranes possess a mixed relaxor and FE domain structure consisting of both in-plane and out-of-plane R polarizations, as well as domains with polarizations along lower symmetry directions, alluding to the presence of either Orthorhombic (O) or monoclinic (M) phases. To aid in understanding the microstructural evolution under electric fields, phase-field simulations on the PMN-PT membrane were performed. The simulations demonstrate that the strain behavior of the PMN-PT membrane can be split into two regions, a low-field region where competing mechanisms result in relatively constant in-plane strain anisotropy as the macroscopic polarization of the membrane switches between up and down, and a high-field region where polarization rotation between $R_{UP}$ and $O_{UP}$ dominates the strain behavior. This work furthers our understanding of the microscopic nature of relaxor-ferroelectric thin films, presenting a significant step towards their use in low-power piezo-driven magnetoelectric devices.

## Results

### Fabrication of (011)-oriented membrane heterostructures

Previously we demonstrated the fabrication of (001)-oriented PMN-PT films on Si(*25*) and fabrication of membranes via etching of a Si substrate.(*33*)  Due to difficulties associated with growth of epitaxial PMN-PT on (011) Si, that method was incompatible for this study. Therefore, we instead use a (011)-oriented $SrTiO_3$ (STO) substrate with a water soluble $Sr_3Al_2O_6$ (SAO) sacrificial layer to create the (011) membrane.(*34*) Fabrication details can be found in the Methods section and key steps are highlighted in Fig. 2. First, epitaxial SAO and a capping STO layer are grown on top of (011)-oriented STO substrates by pulsed laser deposition, followed by sputtering of epitaxial SrRuO₃ (SRO) and PMN-PT layers (Fig. 2A). After depositing a Pt electrode, the entire heterostructure is then attached top-side-down onto a Polydimethylsiloxane (PDMS) and glass platform followed by H₂O etching of the SAO layer to release the films from the STO substrate (Fig. 2B). After removing the STO buffer layer, deposition and patterning of the FM Ni layer into 160μm diameter circular patterns, as well as deposition and patterning of protective polymer Su8 and Au lifted electrode top layers results in the final membrane heterostructure shown in Fig. 2C. A scanning electron microscope (SEM) image of the final heterostructure is shown in Fig. 2D. In order to confirm that the PMN-PT retained its high-quality single-crystalline structure, X-Ray Diffraction (XRD) was performed before and after substrate removal (Fig. S1A). The PMN-PT out-of-plane lattice parameter exhibits no change upon removal of the substrate (Fig. S1B) and the FWHM of the (011) PMN-PT peak rocking curve remains the same as well (Fig. S1C).

## Symmetry-enabled rotation of Ni in-plane anisotropy

Strain-induced changes of magnetic anisotropy in the Ni overlayer were measured by longitudinal magneto-optic Kerr effect (MOKE) hysteresis loops as a function of PMN-PT bias electric fields. When the applied magnetic field is swept along a magnetic easy axis (EA), a square hysteresis loop results from the magnetization reorienting between parallel and antiparallel to the applied field. A magnetic field applied along a hard axis (HA) continuously rotates the magnetization away from the easy axis, resulting in a linear hysteresis loop that saturates at full rotation. Therefore, with field applied along the $[01\bar{1}]_{pc}$ y-direction, we can observe a 90-degree rotation of magnetic anisotropy as the MOKE loop transitions from an EA to HA upon application of the electrical bias, as seen in Fig. 3A.

Due to Ni's negative magnetostriction, it will align its EA along the more compressive direction in the presence of anisotropic strain. At 0 kV/cm bias, the as-grown Ni has a weak EA anisotropy along the y-direction. Application of a large electric bias (positive or negative) results in HA MOKE loops (Fig. 3A, green curves), meaning that the strain is more compressive along the $[100]_{pc}$ x-direction. To confirm that the EA anisotropy is now along the x-direction, MOKE loops were measured with the magnetic field parallel to the $[100]_{pc}$ direction, rotated 90° in-plane from Fig. 3A (Supplemental Fig. S2), and EA MOKE loops were observed at high fields. This matches the expected strain behavior associated with driving the film towards $O_{Up}$ symmetry (Fig. 1). When the electric bias is removed, however, the Ni returns to the as-grown state regardless of bias history, indicating that the strain is relaxed upon removal of the bias. Application of -30 kV/cm bias results in an EA MOKE loop with a higher coercive field (Fig. 3A purple curve). The reinforcement of the EA along the y-direction indicates a reversal in the strain anisotropy from the high-field case, i.e. more compressive along the $[01\bar{1}]_{pc}$ y-direction, and Fig. S2 confirms that the MOKE loop around the FE imprint shows a HA along the x-direction. Therefore, we observe a 90-degree rotation of the Ni in-plane anisotropy over the range of -30 kV/cm to 30 kV/cm applied bias, corresponding to a 3V bias across the thickness of our 500nm PMN-PT membranes. Overall, the MOKE hysteresis behavior is symmetric about -30 kV/cm which we will show to be due to the FE imprint (discussed next section). Similar experiments were performed on a 500nm clamped PMN-30PT thin film still attached to its STO substrate (Fig. S3). Even up to an applied bias of +/-400 kV/cm (+/- 20V) there is no change in the MOKE loop hysteresis. This demonstrates the importance of removing mechanical clamping by the substrate, without which the large anisotropic in-plane strains cannot be achieved.

**Strain behavior and FE properties of PMN-PT/Ni membranes**

To understand the strain behavior inferred from the MOKE hysteresis, we plotted the calculated magnetic anisotropy energy density ($K_U$) determined from the saturation field of hard-axis loops and the associated differential strain ($\varepsilon_{xx} - \varepsilon_{yy}$) using the known magnetostriction of Ni in Fig. 3B. Polarization vs electric field hysteresis loops (PE Loops) are in Fig. 3C and permittivity vs electric field are in Fig. 3D. Supplementary Note 1 details the calculation of $K_U$ and $\varepsilon_{xx} - \varepsilon_{yy}$. The PE loops in Fig. 3C show a FE imprint of approximately -30 kV/cm, which we believe to be due to the asymmetric electrode configuration of SRO (top) and Pt (bottom).(*35*) This results in the zero bias polarization of the PMN-PT film to be in a partially polarized state of ~15 uC/cm$^2$ pointing towards the SRO electrode. The MOKE hysteresis loops (Fig. 3A), as well as the calculated strain (Fig. 3B) and permittivity (Fig. 3D) show similar symmetric behavior about the FE imprint.

In Fig. 3B-D, guidelines have been added that separate the membrane behaviors into three regions: a low-field region near the FE imprint, and high-field regions away from the imprint. In the low-field region, we observe that the strain remains relatively constant (Fig. 3B) while the polarization is switching between negative (down) and positive (up) (Fig. 3C). As will be shown in the next section, the film at 0 kV/cm exhibits a mixture of both in-plane rhombohedral (R$_{IP}$) and out-of-plane rhombohedral (R$_{OP}$) domains. Within the low field region, the film maintains this mixed R state while the polarization switches between positive and negative, resulting in only minor changes to the differential strain. When the polarization begins to saturate in the high-field region, the strain exhibits the largest changes with applied bias as observed in Fig. 3B as well as the MOKE hysteresis in Fig. 3A. The high-field strain behavior arises from monoclinic distortions as the spontaneous polarizations of R out-of-plane domains rotate towards the O direction, as demonstrated later in the phase-field simulation section.

Another interesting feature of the PMN-PT membranes is that the PE loops show a slim-loop hysteresis with low remnant polarization. Similar PE behavior has been reported in other studies of PMN-PT thin films with similar composition(*15, 16, 25, 36*), and resembles the PE hysteresis of a canonical relaxor becoming nonergodic, such as PMN (PMN-xPT with x = 0%) around 250K.(*37*) This supports the claim that the morphotropic phase boundary may be shifted to a higher PT content in PMN-PT thin films (*36*), suggesting that changing the composition of the film may increase the hysteresis of the membrane. The hysteresis also decreases significantly as we approach the DC limit as seen by the quasi-static PE loop at 0.1Hz (see Methods). Since all MOKE measurements must be performed under DC bias, the reduced hysteresis of the quasi-static loop demonstrates that the non-permanent strain behavior observed in MOKE is closely related to the non-permanent polarization behavior of the PMN-PT film.

In the study by Wu et. al. using bulk (011) PMN-PT with a Ni overlayer[4], as well as the study of nano-sized Ni ellipses on bulk (011) PMN-PT by Buzzi et. al. [6], the primary mechanism for anisotropy rotation in the Ni was anisotropic in-plane strain generated by permanent 71°/109° switching between $R_{IP}$ and $R_{UP}$ polarization states. A primarily $R_{IP}$ state would be expected where the overall polarization in the film approaches 0 μC/cm². This would occur near the FE imprint (-30 kV/cm) as seen in Fig. 3C. As we will show in the next section, at 0 kV/cm where the polarization is ~15 μC/cm², we have a mixture of $R_{IP}$ and $R_{OP}$ domains. Therefore, transitioning from the mixed state at 0 kV/cm to a full $R_{IP}$ state at -30 kV/cm would result in $\varepsilon_{xx} - \varepsilon_{yy} < 0$ (Fig. 1), and should result in a HA MOKE loop. But Fig 3A shows an EA MOKE loop at the imprint field instead. Therefore, we do not observe permanent switching to a fully $R_{IP}$ state as observed in bulk (011) PMN-PT.

**STEM analysis of the PMN-PT membrane structure**

The domain structure of the as-grown PMN-PT membranes was investigated via Scanning Transmission Electron Microscope (STEM). Details of the STEM sample and analysis are in the Methods section. A cross-section image of the heterostructure is shown in Fig. S4A. The PMN-PT shows a columnar structure with threading dislocations between the columns that arise due to the lattice mismatch between PMN-PT and SRO/STO during the growth of the PMN-PT film. The selected area diffraction patterns of a single column on each zone axis (Fig. S4B-C) shows that the PMN-PT is single-crystalline. Fig. 4A-B show atomic resolution imaging of the PMN-PT film along the $[01\bar{1}]_{pc}$ and $[100]_{pc}$ zone axes, respectively. To observe the PMN-PT domain structure, we mapped the B-site cation displacement direction and magnitude using the atomic resolution images for both zone axes (Fig. 4C-D). Another representation of the displacements is shown in Fig. 4E-F, where the regions are grouped by corresponding directions for both R and O polarizations for each zone axis (see color wheel next to images).

Since the STEM is performed at zero bias, the FE imprint causes many of the b-site cations to displace either in-plane or down towards the Pt electrode. As seen in Fig. 4C-F, regions of correlated displacements range from less than 1nm up to a few nm in size. These nanoscale domains vary in direction with smooth transitions between them, consisting of both $R_{IP}$ ($R_1$) and $R_{OP}$ ($R_2$) displacements, as well as displacement directions in-between that cannot be classified as either R direction. We label these regions as an orthorhombic $O_1$ and $O_2$, but they could correspond with monoclinic distorted unit cells with polarizations that lie between the two R states as well. Due to the presence of both $R_{IP}$, $R_{OP}$, and polarization states in-between them, we expect the overall differential in-plane strain state at 0 kV/cm to be somewhere between the $R_{UP}$ and $R_{IP}$ states, likely close to the zero strain cubic reference state in Fig. 1C. These findings closely resemble the cation displacement measurements from TEM performed by Kumar et. al.[38] and resemble a mixed FE and relaxor domain structure consistent with the polar slush model.[39]

**Phase-field simulations of PMN-PT membrane**

Phase-field simulations were performed in order to understand the strain behavior of the PMN-PT membrane. Details of the simulation can be found in the Methods section. The initial domain configuration consists of a mixture of $R_{IP}$ and $R_{OP}$ as seen in the spontaneous polarization diagram in Fig. 5A, as well as the [011] (z-direction) stereographic projection of the spontaneous polarizations in Fig. 5B. The evolution of the spontaneous polarization distribution with electric field are shown for 10 kV/cm (Fig. 5C-D), 20 kV/cm (E-F), and 100 kV/cm (G-H). The simulation is summarized by plotting average polarization for the x, y, and z directions (Fig. 5I), as well as the average in-plane strain (Fig. 5J). The average strain was calculated by averaging the strain contribution of individual spontaneous polarization elements multiplied by the electrostriction tensor described in Supplementary Note 1. In the simulation, 0 kV/cm corresponds to no electric bias across the PMN-PT membrane, including any built-in bias from a FE imprint. Therefore, the starting point of the simulation corresponds to the expected structure around the FE imprint of the experimental PMN-PT membrane (-30 kV/cm). Additionally, histograms of the absolute angle between spontaneous polarizations and the $O_{UP}$ [011] direction are shown in Fig. S5.

Guidelines are added to Fig. 5I-J to separate low-field behavior from high-field behavior. As seen in Fig. 5I, the polarization increases rapidly in the low-field region as the mixed $R_{IP}$, $R_{DOWN}$, and $R_{UP}$ state switches to primarily $R_{UP}$ by 20 kV/cm (Fig. 5C-F and S5B-C). Within this region, we observe two behaviors of the spontaneous polarizations within the PMN-PT: 1) Polarization switching from $R_{DOWN}/R_{IP}$ to $R_{UP}$, as seen in the decrease in polarizations near the $R_{IP}$ regions of the stereographic projection, and 2) polarization rotation from $R_{UP}$ towards $O_{UP}$ as indicated by the region between the two $R_{UP}$ polarizations being populated in the stereographic projections. The increase in $R_{UP}$ polarizations and polarization rotation towards $O_{UP}$ (indicated by the shift in the peak towards a lower angle) are also visible in Fig. S5. While the switching between $R_{DOWN}$ and $R_{UP}$ does not result in a change of in-plane strain, the switching between $R_{IP}$ to $R_{UP}$ results in $\varepsilon_{xx} - \varepsilon_{yy} > 0$ (Fig. 1C). On the other hand, the polarization rotation between $R_{UP}$ and $O_{UP}$ results in $\varepsilon_{xx} - \varepsilon_{yy} < 0$. Therefore, the low-field region experiences a large increase in average polarization in the z-direction, while the competing strain behaviors act to keep the in-plane strain relatively constant (Fig. 5I-J).

In the high-field region, the average polarization is nearly saturated but continues to increase as the polarization continues to rotate from $R_{UP}$ towards the $O_{UP}$ state. Nearly all of the polarizations have been switched to $R_{UP}$, meaning that the polarization rotation towards $O_{UP}$ dominates the strain behavior resulting in a large decrease in differential in-plane strain (Fig. 5J). By 100 kV/cm, all of the polarizations lie in the region between $R_{UP}$ and $O_{UP}$ (Fig. 5G-H and S5D), corresponding to a large monoclinic distortion away from $R_{UP}$. The simulation results qualitatively agree with the experimental strain and polarizations measured in the PMN-PT/Ni membrane (Fig. 3), showing relatively constant in-plane strain during polarization switching and large changes of in-plane strain at higher fields. Worth noting is that the strains calculated from the experimental MOKE loops (Fig. 3B) exhibit a horizontal and vertical shift relative to the calculated strains from simulation (Fig. 5J), though qualitatively the two curves are similar. The vertical shift arises from the as-grown Ni being weakly anisotropic with the EA along the y-direction at 0 kV/cm, while the horizontal shift comes from the FE imprint of the PMN-PT membranes being approximately -30 kV/cm.

## Discussion

We have provided initial demonstration of the low-voltage strain-mediated ME effect in an all-thin-film heterostructure that only relies on the large anisotropic strains inherent to (011) PMN-PT thin films by completely removing them from their substrate. The PMN-PT/Ni membranes achieve a robust, piezo-driven, 90° rotation of the in-plane magnetic anisotropy of the Ni overlayer under application of only a few volts bias across the thickness of the PMN-PT membrane. This is roughly two orders of magnitude less voltage than demonstrations using bulk single crystals of PMN-PT which require application of >100V.(*4*) The ME coupling is achieved by driving the PMN-PT polarization towards Orthorhombic symmetry under the applied bias, resulting in strain anisotropy controlled by the in-plane crystal symmetry of the PMN-PT film. STEM measurements show that the zero field domain structure of the PMN-PT membrane consists of a mixture of both $R_{IP}$ and $R_{OP}$ domains, as well as additional regions with *B*-site cation displacements along directions between the two R states. Phase-field simulations confirm that the in-plane differential strain does not change in the low field region near the FE imprint because of the switching between $R_{IP}$ and $R_{OP}$ domains competing with polarization rotation towards $O_{Up}$. However, at higher fields, polarization rotation towards $O_{Up}$ dominates and once again results in a giant piezoelectric effect.

Demonstrations using bulk PMN-PT show permanent switching between in-plane and out-of-plane R polarization states, and consequently between distinct strain states. The permanent switching behavior of bulk PMN-PT is typically a trait that is deemed desirable for applications such as memory storage, but the non-permanent strains in our PMN-PT membranes may still be able to provide the 180 magnetization switching needed for memory devices.(*40, 41*) Understanding the differences between bulk and membrane PMN-PT response to external stimuli is key to their use in future technologies. Such differences may arise due to various effects, such as smaller domain sizes in membrane vs bulk, reduced chemical ordering in PMN-PT membranes due to differences in material processing parameters (e.g. growth temperature), as well as different electrical/mechanical boundary conditions in membranes arising from higher defect concentration, or enhanced role of the interface (*42*). Studying the membrane PMN-PT electrical bias response using *in-situ* STEM or synchrotron x-ray diffraction, as well as studying compositions other than the x = 30% (1-x)PMN-(x)PT used here, could provide key information for future applications. Our work provides key insight to the microstructural behavior of PMN-PT thin film membranes and demonstrates how they can be used in ME coupling devices. Additionally, coupling the PMN-PT membrane with a variety of other materials, such as complex oxides, 2-D materials, and III-V semiconductors, can lead to the discovery of novel piezo-driven phenomena.

## Methods

### Membrane fabrication

20nm of epitaxial $Sr_3Al_2O_6$ (SAO) was grown on top of (011) $SrTiO_3$ (STO) substrates via Pulsed Laser Deposition (PLD). The SAO was grown at a substrate temperature at 780 °C and $p(O_2) = 1 \times 10^{-6}$ Torr using 1 J/cm$^2$ laser fluence on the polycrystalline SAO target. A 20 nm capping layer of STO, preventing possible cation diffusion at the interface[42], was grown at 750 °C and $p(O_2) = 1 \times 10^{-6}$ Torr using 2 J/cm$^2$ laser fluence on the single crystal STO target. 100 nm of SrRuO$_3$ (SRO) was grown by RF magnetron sputtering at a power of 100 W in 200 mTorr of Ar:O$_2$ (12:8) with a substrate temperature of 600°C. 500 nm of 70%Pb(Mg$_{1/3}$Nb$_{2/3}$)O$_3$-30%PbTiO$_3$ (PMN-30%PT) was grown by RF magnetron sputtering at a power of 100 W in 500 mTorr of Ar:O$_2$ (17:3) with a substrate temperature of 625°C. 100 nm of Pt was deposited at room temperature by DC magnetron sputtering. The edges of the heterostructure were ground slightly to remove any side-wall deposition of SRO, PMN-PT, or Pt that would prevent H$_2$O from etching the SAO later on.

Polydimethylsiloxane (PDMS) with a weight ratio of 10:1 (base:crosslinking agent) was spin-coated onto a 10x10 mm$^2$ glass substrate at 5000 rpm for 10 seconds, resulting in a PDMS thickness of ~30 μm. Before the PDMS is cured, the PMN-PT film heterostructure is placed into the PDMS Pt-side down. The entire sample is placed in vacuum for a minimum of 5 hours to remove any bubbles between the PDMS and Pt layers, followed by curing of the PDMS on a hot plate at 100°C for 1 hour. Placing the film into the uncured PDMS is crucial, as it allows the PDMS to mold to the surface of the film ensuring that the PMN-PT membrane is as flat as possible after the substrate is removed. One consequence of curing the PDMS after the film is attached is that it will mold to the sides of the substrate. Once cured, removing the PDMS from the sides of the substrate with a razor is necessary so that the H$_2$O can reach the SAO.

The sample is placed in a beaker of water in order to etch the sacrificial SAO layer. This process can take anywhere from 24-72 hours. Heating the water to 70-80 °C was found to speed up the etching in some cases. Etching progress was monitored by visual inspection under a microscope, and once found to be completed, the substrate was removed using tweezers. Dipping in isopropyl alcohol was used to displace water and reduce surface tension between the substrate and film in some instances.

With the substrate removed, the exposed surface of the membrane now consists of the STO layer that was used to cap the SAO layer. Ion-milling was used to remove the STO layer and expose the SRO film. A 35 nm Ni film was deposited by DC sputtering at room temperature on top of the SRO to act as the FM layer for our FE/FM composite. Photolithography and wet etching were used to pattern the SRO/Ni into 160 μm diameter disks. Su8 photoresist was spin-coated and patterned by photolithography to create a protection layer that left the Ni/SRO disks exposed while covering the PMN-PT. 30 nm of Au was deposited by DC sputtering at room temperature, followed by photolithography and patterning to create Au electrodes partially overlapping the Ni/SRO disks, and partially on top of the Su8. This allowed for electrical contact to be made with probe tips or wire-bonding to the Au on the Su8 layer without risk of damaging the fragile membrane heterostructure.

**Scanning transmission electron microscopy**

Two kinds of cross-sectional samples having [100] and [01$\bar{1}$] pseudocubic projections were prepared using a dual-beam focused ion beam system (Helios G3, FEI) to determine the ferroelectric domain structure. We used a Ga ion beam at 30 kV to make a thin specimen and then used different acceleration voltages from 5 to 1 kV for the sample cleaning process to reduce the Ga damage. The selected area diffraction pattern analysis and atomic structure observation were performed using a STEM (JEM-ARM200F, JEOL, Japan) at 200 kV equipped with a fifth-order probe corrector (ASCOR, CEOS GmbH, Germany) at Materials Imaging & Analysis Center of POSTECH in South Korea. The optimum size of the electron probe for STEM observation was ~78 pm. The collection semiangles of the HAADF detector were adjusted from 68 to 280 mrad to collect scattered electrons in a large angle for clear Z-sensitive images. HAADF-STEM images were acquired using Smart Align (HREM Research Inc., Japan), which conducted the multi-stack of images and aligned them using rigid registration to correct the sample drift and scan distortions. The obtained raw images were processed using a bandpass difference filter with a local window to reduce background noise (Filters Pro, HREM Research Inc., Japan).

STEM image analysis was performed by MATLAB with the customized atomic analysis code. All atomic coordinates were determined by the centroid of each atomic column. Chemical classification on A and B site was conducted with Z-contrast difference on HAADF-STEM image and unit cell was defined as B-site with neighboring 4 A-sites in $[100]_{pc}$ projection and B site with neighboring 6 A-sites in $[01\bar{1}]_{pc}$ projection. B-site off centering was defined by the displacement of B-site position from the mean neighbor positions of A-sites for each unit cell.

**Phase-field simulations**

The phase-field method is employed to simulate the effect of applied bias on the polarization distribution in (011) PMN-30PT freestanding membranes. In the phase-field model, the polarization is selected as the order parameter to describe the domain structures, and its spatial and temporal evolutions is controlled by the time-dependent Ginzburg-Landau equation(*43, 44*): $(\partial P_i/\partial t) = -L(\partial F/\partial P_i)$ with $L$ the kinetic coefficient related to the domain-wall mobility and $F$ the total free energy which includes the bulk chemical energy, polarization gradient energy, electric energy, and elastic energy.(*43*) The parameters for the bulk chemical energy is from literature.(*45*) The freestanding membrane is represented by a grid of $128\Delta x \times 128\Delta x \times (N_{\text{bottom-air}} + N_{\text{membrane}} + N_{\text{top-air}})\Delta z$ with $\Delta x = \Delta z = 1$ nm, $N_{\text{bottom-air}} = 2$, $N_{\text{membrane}} = 20$, and $N_{\text{top-air}} = 2$. Periodic boundary conditions are assumed in the in-plane directions. To consider the (011) orientation, we define the simulation coordinate system (*x, y, z*) to be x//[100], y//[01$\bar{1}$] z//[011], and the tensor coefficients are rotated by the corresponding rotation matrix.

Stress-free boundary conditions are assumed for both the top and bottom surfaces for solving the elastic equilibrium equation of the membrane. This is achieved by considering an inhomogeneous system including two layers of vacuum at both the top and bottom surfaces of the membrane. The elastic equilibrium equation of such an inhomogeneous system is solved using the spectral iterative perturbation method.(*46*) To incorporate the applied electric field, a superposition method is employed to solve the electrostatic equilibrium equation with a uniform preset voltage bias at the top surface while the bottom surface is grounded.(*47*)

**Magneto-optic Kerr effect measurements.**

The PMN-PT membrane was mounted between the poles of an electromagnet and a polarized red HeNe (632 nm) laser was reflected off of the sample surface at approximately 45° from normal incidence. The beam was focused to an approximately 10 μm spot near the center of the 160 μm Ni discs using an achromat lens. The reflected beam's polarization was analyzed with a differential detection scheme. A polarizing beam splitter directed the *s-* and *p-* components of the reflected beam onto two channels of a Thorlabs PDB210A differential photodetector, and a half-wave plate before the beam splitter was used to balance the inputs to the detector. A fitting procedure was used to extract $H_{sat}$ from hard axis MOKE loops, resulting in the values plotted in Fig. 4d. The loop was normalized so that the Kerr rotation at magnetic saturation is $\pm 1$. Data with normalized Kerr rotation values between 0.85 and -0.85 were fit to a line. The magnetic field values where the fitted line intersects with +1 and -1 normalized Kerr rotation were respectively taken to be $H_{sat}^+$ and $H_{sat}^-$, with $H_{sat}$ the average.

### Ferroelectric measurements.

Permittivity measurements were performed by slowly (0.5 Hz) sweeping a bias electric field across a device while also applying an AC waveform (3.5 kV/cm RMS at 4 kHz), detecting the resulting 4 kHz AC current with a lock-in amplifier. High frequency P-E loops were obtained by integrating the current flowing to the device while sweeping the electric field at 30 kHz. The observed frequency dependence of the P-E loop required determining the P-E response at very low frequency in order to compare with MOKE measurements, where magnetic hysteresis data were acquired with static electric field. An 0.1Hz equivalent polarization at an electric field $E_0$ was determined by letting the polarization decay from its initial high-frequency sweep value for 10s at fixed electric field $E_0$, then saturating with a high frequency sweep and integrating the resulting current to determine the polarization change. Both the initializing sweep and saturating sweep were complete PE loops with the same electric field extent, but phase shifted to begin and end at $E_0$.

# References

1. J.-M. Hu, L.-Q. Chen, C.-W. Nan, Multiferroic Heterostructures Integrating Ferroelectric and Magnetic Materials. *Adv. Mater.* **28**, 15–39 (2016).

2. Z. Chu, M. PourhosseiniAsl, S. Dong, Review of multi-layered magnetoelectric composite materials and devices applications. *J. Phys. Appl. Phys.* **51**, 243001 (2018).

3. S. Zhang, F. Li, High performance ferroelectric relaxor-$PbTiO_3$ single crystals: Status and perspective. *J. Appl. Phys.* **111**, 031301 (2012).

4. T. Wu, A. Bur, P. Zhao, K. P. Mohanchandra, K. Wong, K. L. Wang, C. S. Lynch, G. P. Carman, Giant electric-field-induced reversible and permanent magnetization reorientation on magnetoelectric $Ni/(011)[Pb(Mg_{1/3}Nb_{2/3})O_3]_{(1-x)}–[PbTiO_3]_x$ heterostructure. *Appl. Phys. Lett.* **98**, 012504 (2011).

5. Z. Wang, Y. Wang, H. Luo, J. Li, D. Viehland, Crafting the strain state in epitaxial thin films: A case study of $CoFe_2O_4$ films on $Pb(Mg,Nb)O_3–PbTiO_3$. *Phys. Rev. B.* **90**, 134103 (2014).

6. M. Buzzi, R. V. Chopdekar, J. L. Hockel, A. Bur, T. Wu, N. Pilet, P. Warnicke, G. P. Carman, L. J. Heyderman, F. Nolting, Single Domain Spin Manipulation by Electric Fields in Strain Coupled Artificial Multiferroic Nanostructures. *Phys. Rev. Lett.* **111**, 027204 (2013).

7. A. Chen, Y. Wen, B. Fang, Y. Zhao, Q. Zhang, Y. Chang, P. Li, H. Wu, H. Huang, Y. Lu, Z. Zeng, J. Cai, X. Han, T. Wu, X.-X. Zhang, Y. Zhao, Giant nonvolatile manipulation of magnetoresistance in magnetic tunnel junctions by electric fields via magnetoelectric coupling. *Nat. Commun.* **10**, 243 (2019).

8. M. Liu, B. M. Howe, L. Grazulis, K. Mahalingam, T. Nan, N. X. Sun, G. J. Brown, Voltage-Impulse-Induced Non-Volatile Ferroelastic Switching of Ferromagnetic Resonance for Reconfigurable Magnetoelectric Microwave Devices. *Adv. Mater.* **25**, 4886–4892 (2013).

9. T. Nan, M. Liu, W. Ren, Z.-G. Ye, N. X. Sun, Voltage Control of Metal-insulator Transition and Non-volatile Ferroelastic Switching of Resistance in $VO_x$/PMN-PT Heterostructures. *Sci. Rep.* **4**, 5931 (2015).

10. B. Noheda, D. E. Cox, G. Shirane, J. Gao, Z.-G. Ye, Phase diagram of the ferroelectric relaxor $(1-x)PbMg_{1/3}Nb_{2/3}O_3–xPbTiO_3$. *Phys. Rev. B.* **66**, 054104 (2002).

11. M. Davis, Picturing the elephant: Giant piezoelectric activity and the monoclinic phases of relaxor-ferroelectric single crystals. *J. Electroceramics.* **19**, 25–47 (2007).

12. F. Li, L. Jin, Z. Xu, D. Wang, S. Zhang, Electrostrictive effect in $Pb(Mg_{1/3}Nb_{2/3})O_3$-$xPbTiO_3$ crystals. *Appl. Phys. Lett.* **102**, 152910 (2013).

13. J.-M. Hu, Z. Li, L.-Q. Chen, C.-W. Nan, High-density magnetoresistive random access memory operating at ultralow voltage at room temperature. *Nat. Commun.* **2**, 553 (2011).

14. S. Manipatruni, D. E. Nikonov, C.-C. Lin, T. A. Gosavi, H. Liu, B. Prasad, Y.-L. Huang, E. Bonturim, R. Ramesh, I. A. Young, Scalable energy-efficient magnetoelectric spin–orbit logic. *Nature.* **565**, 35–42 (2019).

15. M. Boota, E. P. Houwman, M. Dekkers, M. D. Nguyen, K. H. Vergeer, G. Lanzara, G. Koster, G. Rijnders, Properties of epitaxial, (001)- and (110)-oriented $(PbMg_{1/3}Nb_{2/3}O_3)_{2/3}$- (PbTiO$_3$)$_{1/3}$ films on silicon described by polarization rotation. *Sci. Technol. Adv. Mater.* **17**, 45–57 (2016).

16. R. Keech, C. Morandi, M. Wallace, G. Esteves, L. Denis, J. Guerrier, R. L. Johnson-Wilke, C. M. Fancher, J. L. Jones, S. Trolier-McKinstry, Thickness-dependent domain wall reorientation in 70/30 lead magnesium niobate-lead titanate thin films. *J. Am. Ceram. Soc.* **100**, 3961–3972 (2017).

17. S. Pandya, J. Wilbur, J. Kim, R. Gao, A. Dasgupta, C. Dames, L. W. Martin, Pyroelectric energy conversion with large energy and power density in relaxor ferroelectric thin films. *Nat. Mater.* **17**, 432–438 (2018).

18. K. Lefki, G. J. M. Dormans, Measurement of piezoelectric coefficients of ferroelectric thin films. *J. Appl. Phys.* **76**, 1764–1767 (1994).

19. S. Trolier-McKinstry, P. Muralt, Thin Film Piezoelectrics for MEMS. *J. Electroceramics*. **12**, 7–17 (2004).

20. P. B. Meisenheimer, S. Novakov, N. M. Vu, J. T. Heron, Perspective: Magnetoelectric switching in thin film multiferroic heterostructures. *J. Appl. Phys.* **123**, 240901 (2018).

21. G. P. Carman, N. Sun, Strain-mediated magnetoelectrics: Turning science fiction into reality. *MRS Bull.* **43**, 822–828 (2018).

22. H. Palneedi, V. Annapureddy, S. Priya, J. Ryu, Status and Perspectives of Multiferroic Magnetoelectric Composite Materials and Applications. *Actuators*. **5**, 9 (2016).

23. J. Ma, J. Hu, Z. Li, C.-W. Nan, Recent Progress in Multiferroic Magnetoelectric Composites: from Bulk to Thin Films. *Adv. Mater.* **23**, 1062–1087 (2011).

24. W. Gao, Y. Zhu, Y. Wang, G. Yuan, J.-M. Liu, A review of flexible perovskite oxide ferroelectric films and their application. *J. Materiomics*. **6**, 1–16 (2020).

25. S. H. Baek, J. Park, D. M. Kim, V. A. Aksyuk, R. R. Das, S. D. Bu, D. A. Felker, J. Lettieri, V. Vaithyanathan, S. S. N. Bharadwaja, N. Bassiri-Gharb, J. B. Chen, H. P. Sun, C. M. Folkman, H. W. Jang, D. J. Kreft, S. K. Streiffer, R. Ramesh, X. Q. Pan, S. Trolier-McKinstry, D. G. Schlom, M. S. Rzchowski, R. H. Blick, C. B. Eom, Giant Piezoelectricity on Si for Hyperactive MEMS. *Science*. **334**, 958–961 (2011).

26. R. Keech, L. Ye, J. L. Bosse, G. Esteves, J. Guerrier, J. L. Jones, M. A. Kuroda, B. D. Huey, S. Trolier-McKinstry, Declamped Piezoelectric Coefficients in Patterned 70/30 Lead Magnesium Niobate-Lead Titanate Thin Films. *Adv. Funct. Mater.* **27**, 1605014 (2017).

27. V. Nagarajan, A. Roytburd, A. Stanishevsky, S. Prasertchoung, T. Zhao, L. Chen, J. Melngailis, O. Auciello, R. Ramesh, Dynamics of ferroelastic domains in ferroelectric thin films. *Nat. Mater.* **2**, 43–47 (2003).

28. S. Bühlmann, B. Dwir, J. Baborowski, P. Muralt, Size effect in mesoscopic epitaxial ferroelectric structures: Increase of piezoelectric response with decreasing feature size. *Appl. Phys. Lett.* **80**, 3195–3197 (2002).

29. M. Wallace, R. L. Johnson-Wilke, G. Esteves, C. M. Fancher, R. H. T. Wilke, J. L. Jones, S. Trolier-McKinstry, In situ measurement of increased ferroelectric/ferroelastic domain wall motion in declamped tetragonal lead zirconate titanate thin films. *J. Appl. Phys.* **117**, 054103 (2015).

30. F. Griggio, S. Jesse, A. Kumar, O. Ovchinnikov, H. Kim, T. N. Jackson, D. Damjanovic, S. V. Kalinin, S. Trolier-McKinstry, Substrate Clamping Effects on Irreversible Domain Wall Dynamics in Lead Zirconate Titanate Thin Films. *Phys. Rev. Lett.* **108**, 157604 (2012).

31. H. Greve, E. Woltermann, H.-J. Quenzer, B. Wagner, E. Quandt, Giant magnetoelectric coefficients in $(Fe_{90}Co_{10})_{78}Si_{12}B_{10}$-AlN thin film composites. *Appl. Phys. Lett.* **96**, 182501 (2010).

32. D. Viehland, M. Wuttig, J. McCord, E. Quandt, Magnetoelectric magnetic field sensors. *MRS Bull.* **43**, 834–840 (2018).

33. J. Irwin, S. Lindemann, W. Maeng, J. J. Wang, V. Vaithyanathan, J. M. Hu, L. Q. Chen, D. G. Schlom, C. B. Eom, M. S. Rzchowski, Magnetoelectric Coupling by Piezoelectric Tensor Design. *Sci. Rep.* **9**, 19158 (2019).

34. D. Lu, D. J. Baek, S. S. Hong, L. F. Kourkoutis, Y. Hikita, H. Y. Hwang, Synthesis of freestanding single-crystal perovskite films and heterostructures by etching of sacrificial water-soluble layers. *Nat. Mater.* **15**, 1255–1260 (2016).

35. J. Lee, C. H. Choi, B. H. Park, T. W. Noh, J. K. Lee, Built-in voltages and asymmetric polarization switching in Pb(Zr,Ti)O₃ thin film capacitors. *Appl. Phys. Lett.* **72**, 3380–3382 (1998).

36. S. Yokoyama, S. Okamoto, H. Funakubo, T. Iijima, K. Saito, H. Okino, T. Yamamoto, K. Nishida, T. Katoda, J. Sakai, Crystal structure, electrical properties, and mechanical response of $(100)$-/$(001)$-oriented epitaxial $Pb(Mg_{1/3}Nb_{2/3})O_3$–$PbTiO_3$ films grown on $(100)_c$SrRuO₃|$(100)$SrTiO₃ substrates by metal-organic chemical vapor deposition. *J. Appl. Phys.* **100**, 054110 (2006).

37. D. Fu, H. Taniguchi, M. Itoh, S. Koshihara, N. Yamamoto, S. Mori, Relaxor $Pb(Mg_{1/3}Nb_{2/3})O_3$: A Ferroelectric with Multiple Inhomogeneities. *Phys. Rev. Lett.* **103**, 207601 (2009).

38. A. Kumar, J. N. Baker, P. C. Bowes, M. J. Cabral, S. Zhang, E. C. Dickey, D. L. Irving, J. M. LeBeau, Atomic-resolution electron microscopy of nanoscale local structure in lead-based relaxor ferroelectrics. *Nat. Mater.* **20**, 62-67 (2021).

39. H. Takenaka, I. Grinberg, S. Liu, A. M. Rappe, Slush-like polar structures in single-crystal relaxors. *Nature.* **546**, 391–395 (2017).

40. J.-M. Hu, T. Yang, J. Wang, H. Huang, J. Zhang, L.-Q. Chen, C.-W. Nan, Purely Electric-Field-Driven Perpendicular Magnetization Reversal. *Nano Lett.* **15**, 616–622 (2015).

41. J.-M. Hu, T. Yang, K. Momeni, X. Cheng, L. Chen, S. Lei, S. Zhang, S. Trolier-McKinstry, V. Gopalan, G. P. Carman, C.-W. Nan, L.-Q. Chen, Fast Magnetic Domain-Wall Motion in a Ring-Shaped Nanowire Driven by a Voltage. *Nano Lett.* **16**, 2341–2348 (2016).

42. M.G. Han, M. S.J. Marshall, L. Wu, M. A. Schofield, T. Aoki, R. Twesten, J. Hoffman, F. J. Walker, C. H. Ahn & Y. Zhu, Interface-Induced Nonswitchable Domains in Ferroelectric Thin Films. *Nat. Commun.* **5,** 4693 (2014).

43. J.-J. Wang, B. Wang, L.-Q. Chen, Understanding, Predicting, and Designing Ferroelectric Domain Structures and Switching Guided by the Phase-Field Method. *Annu. Rev. Mater. Res.* **49**, 127–152 (2019).

44. L.-Q. Chen, Phase-Field Method of Phase Transitions/Domain Structures in Ferroelectric Thin Films: A Review. *J. Am. Ceram. Soc.* **91**, 1835–1844 (2008).

45. H. Zhang, X. Lu, R. Wang, C. Wang, L. Zheng, Z. Liu, C. Yang, R. Zhang, B. Yang, W. Cao, Phase coexistence and Landau expansion parameters for a $0.70Pb(Mg_{1/3}Nb_{2/3})O_3-0.30PbTiO_3$ single crystal. *Phys. Rev. B*. **96**, 054109 (2017).

46. J. J. Wang, X. Q. Ma, Q. Li, J. Britson, L.-Q. Chen, Phase transitions and domain structures of ferroelectric nanoparticles: Phase field model incorporating strong elastic and dielectric inhomogeneity. *Acta Mater.* **61**, 7591–7603 (2013).

47. Y. L. Li, S. Y. Hu, Z. K. Liu, L. Q. Chen, Effect of electrical boundary conditions on ferroelectric domain structures in thin films. *Appl. Phys. Lett.* **81**, 427–429 (2002).

48. J. F. Nye, *Physical properties of crystals: their representation by tensors and matrices* (Oxford University Press, Oxford, England, 1957), pp. 134.

49. M. J. Haun, E. Furman, S. J. Jang, L. E. Cross, Modeling of the electrostrictive, dielectric, and piezoelectric properties of ceramic $PbTiO_3$. *IEEE Trans. Ultrason. Ferroelectr. Freq. Control*. **36**, 393–401 (1989).

50. R. E. Newnham, V. Sundar, R. Yimnirun, J. Su, Q. M. Zhang, Electrostriction: Nonlinear Electromechanical Coupling in Solid Dielectrics. *J. Phys. Chem. B*. **101**, 10141–10150 (1997).

51. C. Tannous, J. Gieraltowski, The Stoner–Wohlfarth model of ferromagnetism. *Eur. J. Phys.* **29**, 475–487 (2008).

52. C. Kittel, Physical Theory of Ferromagnetic Domains. *Rev. Mod. Phys.* **21**, 541–583 (1949).

**Acknowledgement**

**Funding:** This work was supported by the Army Research Office through Grant W911NF-17-1-0462 and Vannevar Bush Faculty Fellowship (N00014-20-1-2844) and the Gordon and Betty Moore Foundation's EPiQS Initiative, grant GBMF9065 to C.B.E. MOKE measurement and analytical strain calculations at the University of Wisconsin–Madison was supported by the US Department of Energy (DOE), Office of Science, Office of Basic Energy Sciences (BES), under award number DE-FG02-06ER46327. S.Y.C. acknowledges the support of the Global Frontier Hybrid Interface Materials of the National Research Foundation of Korea (NRF) funded by the Ministry of Science and ICT (2013M3A6B1078872).

**Author contributions:** M.S.R., C.B.E. and S.Y.C. supervised the experiments. L.Q.C. supervised theoretical calculations. S.L. and K.E. fabricated and characterized thin film membrane heterostructures. J.I. performed analytical strain calculations. S.L. and J.I. carried out magnetoelectric coupling experiments. G.Y.K. and S.Y.C. carried out scanning transmission electron microscopy and polarization displacement mapping analysis. B.W. and J.J.W. performed theoretical calculations. J.M.H. contributed to interpretation of results. M.S.R. and C.B.E. directed the overall research.

**Competing interests:** The authors declare that they have no competing interests.

**Data and materials availability:** All data needed to evaluate the conclusions in the paper are present in the paper and/or the Supplementary Materials.

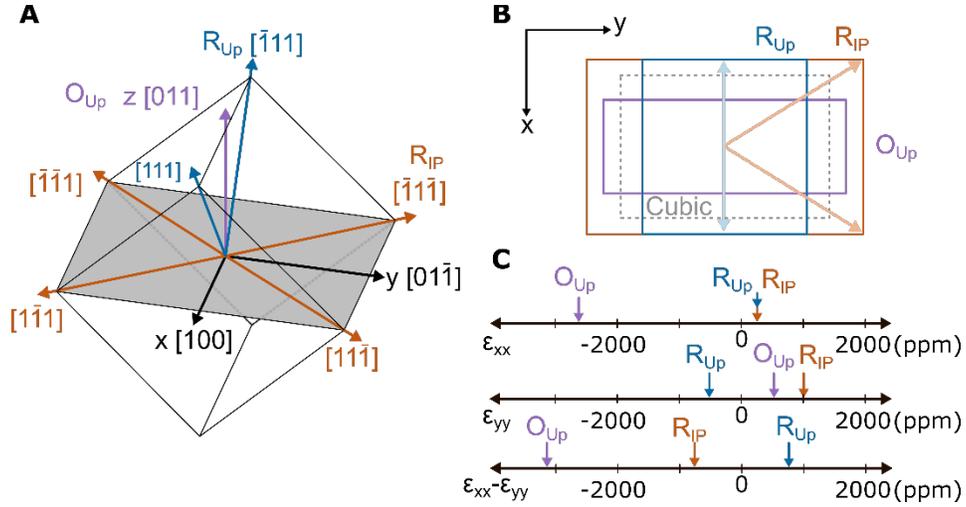

**Fig. 1. Anisotropic strain in (011)-oriented PMN-PT. (A)** Cartesian coordinates $x$, $y$ and $z$ are defined to be the crystal [100], [01$\bar{1}$], and [011] directions. Polarization directions in (011)-oriented PMN-PT unit cell, grouped into rhombohedral in-plane ($R_{IP}$, orange), rhombohedral up ($R_{Up}$, blue) and orthorhombic up ($O_{Up}$, purple). Rhombohedral down ($R_{Down}$) and orthorhombic down ($O_{Down}$) are not shown but are respectively $R_{Up}$ and $O_{Up}$ mirrored about the $xy$-plane. The in-plane cut through the unit cell (shaded grey area) is rectangular with sides of length $a\sqrt{2}$ by $a$, where $a$ is the lattice parameter. **(B)** Electrostrictive deformations (not to scale) of the unit cell for the cubic (zero ferroelectric polarization), $R_{IP}$, $R_{Up}$ and $O_{Up}$ polarization groups. The down deformations are identical to up. In-plane projections of polarization vectors are shown for $R_{IP}$ (light orange) and $R_{Up}$ (light blue). **(C)** Plots of linear electrostriction strains $\varepsilon_{xx}$ and $\varepsilon_{yy}$ and the anisotropic strain $\varepsilon_{xx} - \varepsilon_{yy}$ for $R_{IP}$, $R_{Up}$ and $O_{Up}$ polarization groups. Numerical values are in Supplementary Table 1.

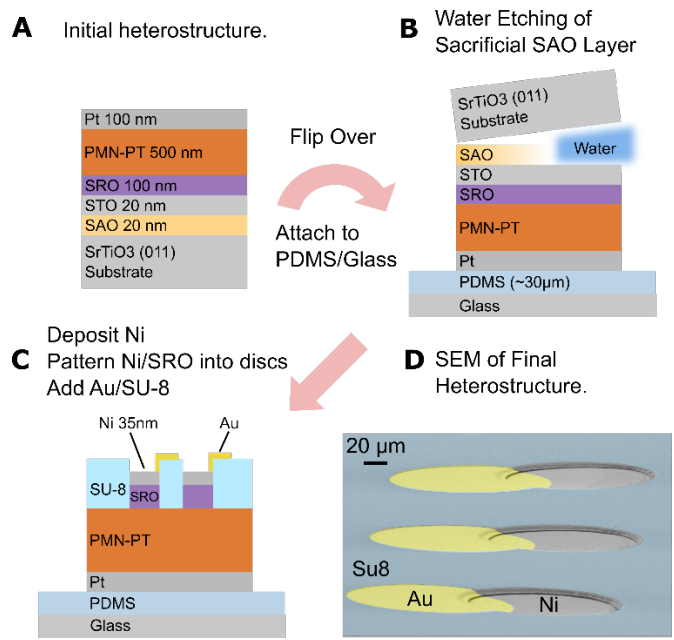

**Fig. 2. Fabrication of single-crystal (011)-oriented PMN-PT membrane heterostructures. (A)** Initial thin film heterostructure consisting of PLD-grown SAO/STO layers and sputter deposited SRO/PMN-PT/Pt layers. **(B)** After attaching the heterostructure Pt-side into PDMS/Glass, the SAO sacrificial layer is etched by $H_2O$. **(C)** After removal of the STO buffer layer, Ni is deposited by sputtering followed by patterning of the Ni/SRO layers into 160μm circles. The membrane heterostructure is completed by addition of the Su8 protective layer and Au lifted electrode layer. **(D)** SEM image showing the completed membrane device.

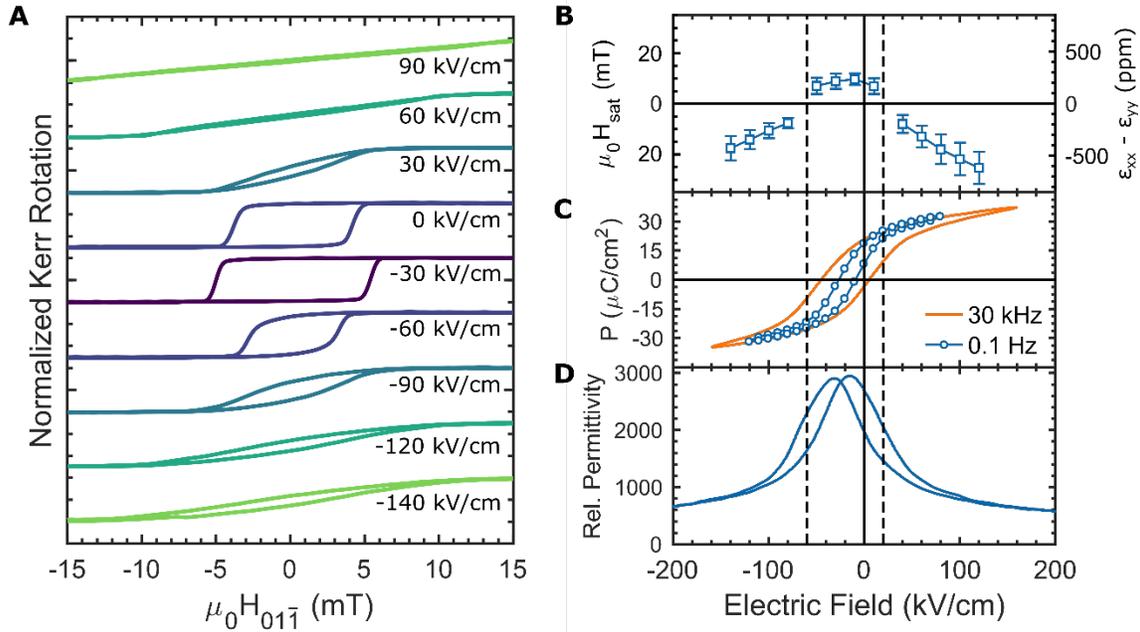

**Fig. 3. Magnetoelectric, ferroelectric and piezoelectric properties of PMN-PT membrane heterostructures. (A)** MOKE magnetic hysteresis loops (normalized) at a series of electric fields from -140 kV/cm (-7 V) to 90 kV/cm (4.5 V). Dark colors are closer to the ferroelectric imprint and lighter colors are further from the imprint. **(B)** Saturation magnetic field ($H_{sat}$, left axis) and calculated anisotropic strain ($\varepsilon_{xx} - \varepsilon_{yy}$, right axis) versus biasing electric field extracted from hard axis MOKE hysteresis loops similar to those shown at high bias electric field in (A). Error bars represent standard deviation of measurements of seven different devices on the same membrane. Negative differential strain points ($\varepsilon_{xx} - \varepsilon_{yy} < 0$) were extracted from hard axis MOKE loops with magnetic field along [01$\bar{1}$], and positive points ($\varepsilon_{xx} - \varepsilon_{yy} > 0$) from loops where magnetic field was along [100]. **(C)** P-E loop measurements using 160 µm diameter Ni/SrRuO$_3$ top electrode. The orange loop was measured with a 30 kHz sinusoidal voltage pulse. The blue curve, labelled as 0.1 Hz, was acquired using a quasi-DC measurement procedure (see Methods). **(D)** Relative permittivity versus biasing electric field. Bias electric field was swept at 0.5 Hz, and permittivity was measured with a small AC electric field of 3.5 kV/cm RMS at 4 kHz. For **(B)-(D)**, guidelines are added to separate the behavior into a low field region (near FE imprint) and high-field regions.

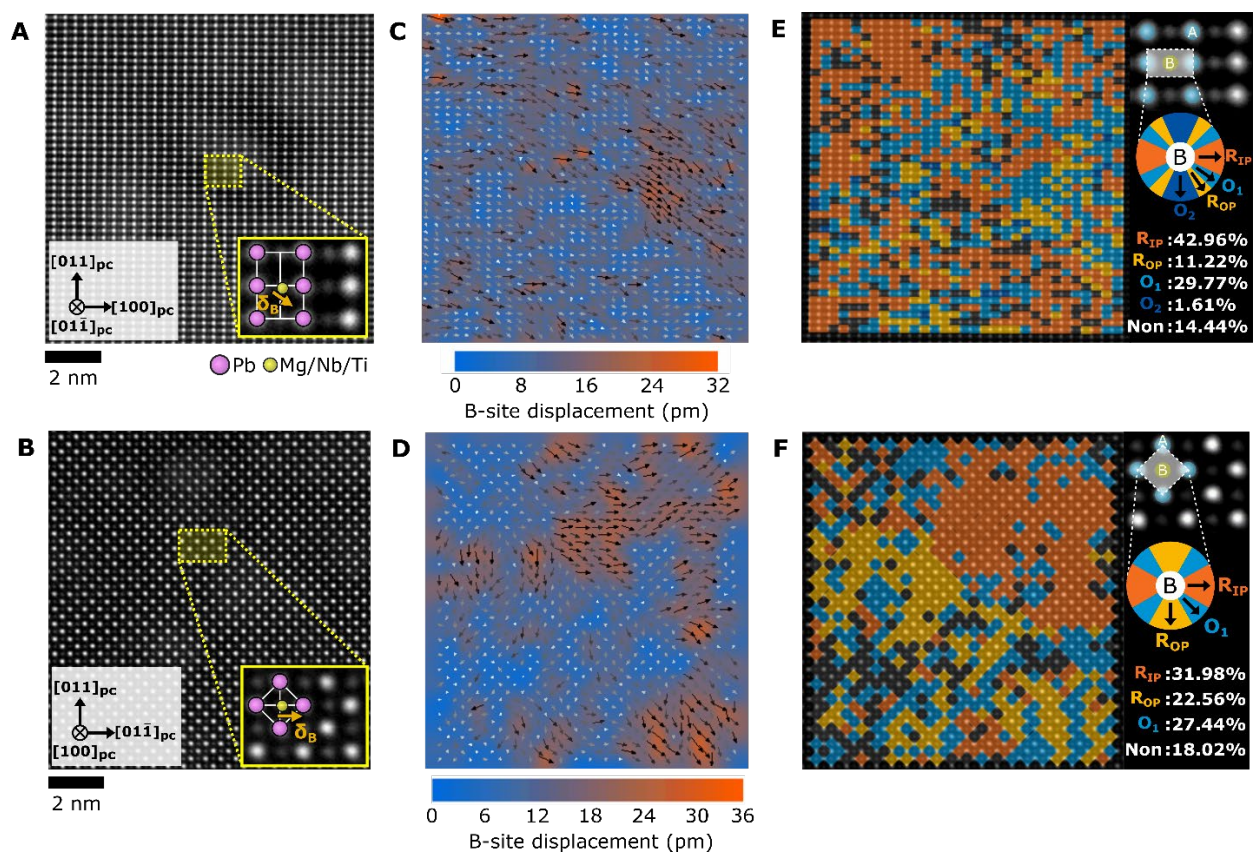

**Fig. 4. STEM analysis of domains present in PMN-PT membrane. (A-B)** Atomic resolution HAADF-STEM images along the $[01\bar{1}]_{pc}$ and $[100]_{pc}$ zone axes, respectively. The insets are enlarged images in each zone axis. Pink circles are A-site cations (Pb) and yellow circles are B-site cations (Mg/Nb/Ti). Orange arrows are the b-site displacement ($\delta_b$). **(C-D)** B-site cation displacement mapping with overlaid arrows indicating regions of short-range ordering. Color maps show the atomic displacement magnitude, and arrows display the direction of atomic displacement. **(E-F)** Phase fraction mapping in each unit-cell with color wheel by expected B-site displacement directions for $R_{IP}$ ($R_1$), $R_{OP}$ ($R_2$), and regions that have displacements between the R states labeled as orthorhombic $O_1$ and $O_2$. Color blank regions (Non) indicate the non-polar region under the 7 picometers of B-site displacement.

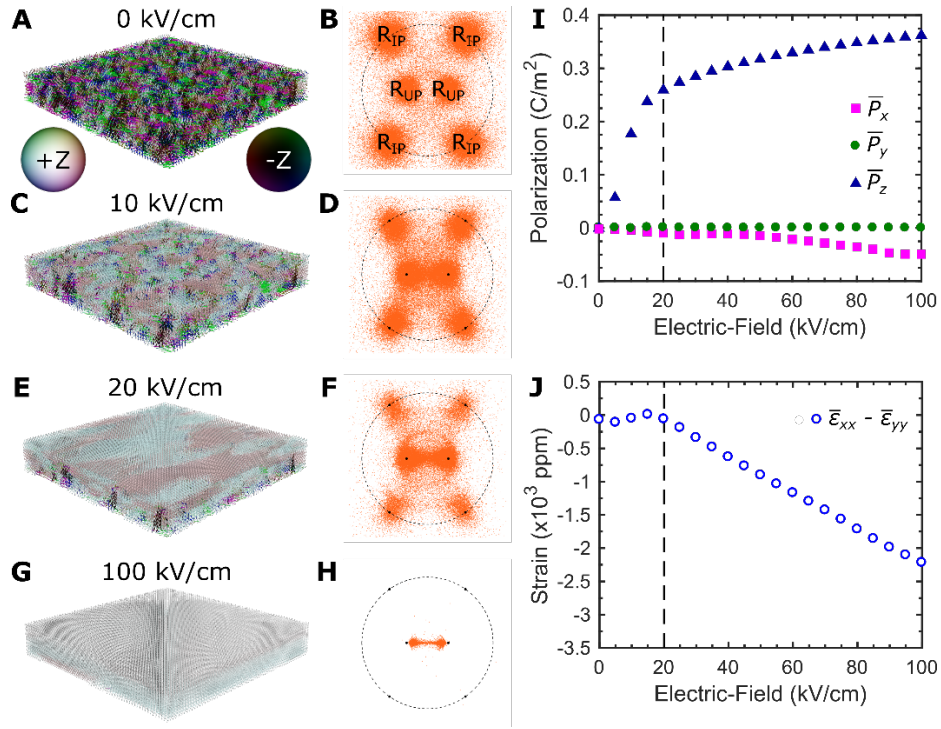

**Fig. 5. Phase-field simulations of (011) PMN-PT membrane.** Spontaneous polarization and [011] stereographic projection of the PMN-PT membrane at **(A)-(B)** 0 kV/cm, **(C)-(D)** 10 kV/cm, **(E)-(F)** 20 kV/cm, and **(G)-(H)** 100 kV/cm. The legend for the coloring of spontaneous polarization is included in (A). **(I)** Average polarization in the x, y, and z directions vs applied field. **(J)** Field dependence of the average anisotropic in-plane strain $\bar{\varepsilon}_{xx} - \bar{\varepsilon}_{yy}$. In (I)-(J), guidelines have been added to separate the low-field and high-field regions.

# Supporting Information for:

## Low-Voltage Magnetoelectric Coupling in Membrane Heterostructures

S. Lindemann,[1,†] J. Irwin,[2,†] G.-Y. Kim,[3] B. Wang,[4] K. Eom,[1] J. J. Wang,[4] J. M. Hu,[1] L. Q. Chen,[4] S. Y. Choi,[3] C. B. Eom,[1*] M. S. Rzchowski,[2,*]

[1]Department of Materials Science and Engineering, University of Wisconsin-Madison
Madison, Wisconsin 53706, United States.

[2] Department of Physics, University of Wisconsin-Madison
Madison, Wisconsin 53706, United States.

[3] Department of Materisals Science and Engineering, Pohang University of Science and Technology, Pohang, Gyeongbuk 37673, Korea.

[4]Department of Materials Science and Engineering, Pennsylvania State University, University Park, Pennsylvania 16802, United States.

[†]J.I. and S.L. contributed equally to this work

[*]To whom correspondence should be addressed. E-mail: eom@engr.wisc.edu, rzchowski@physics.wisc.edu

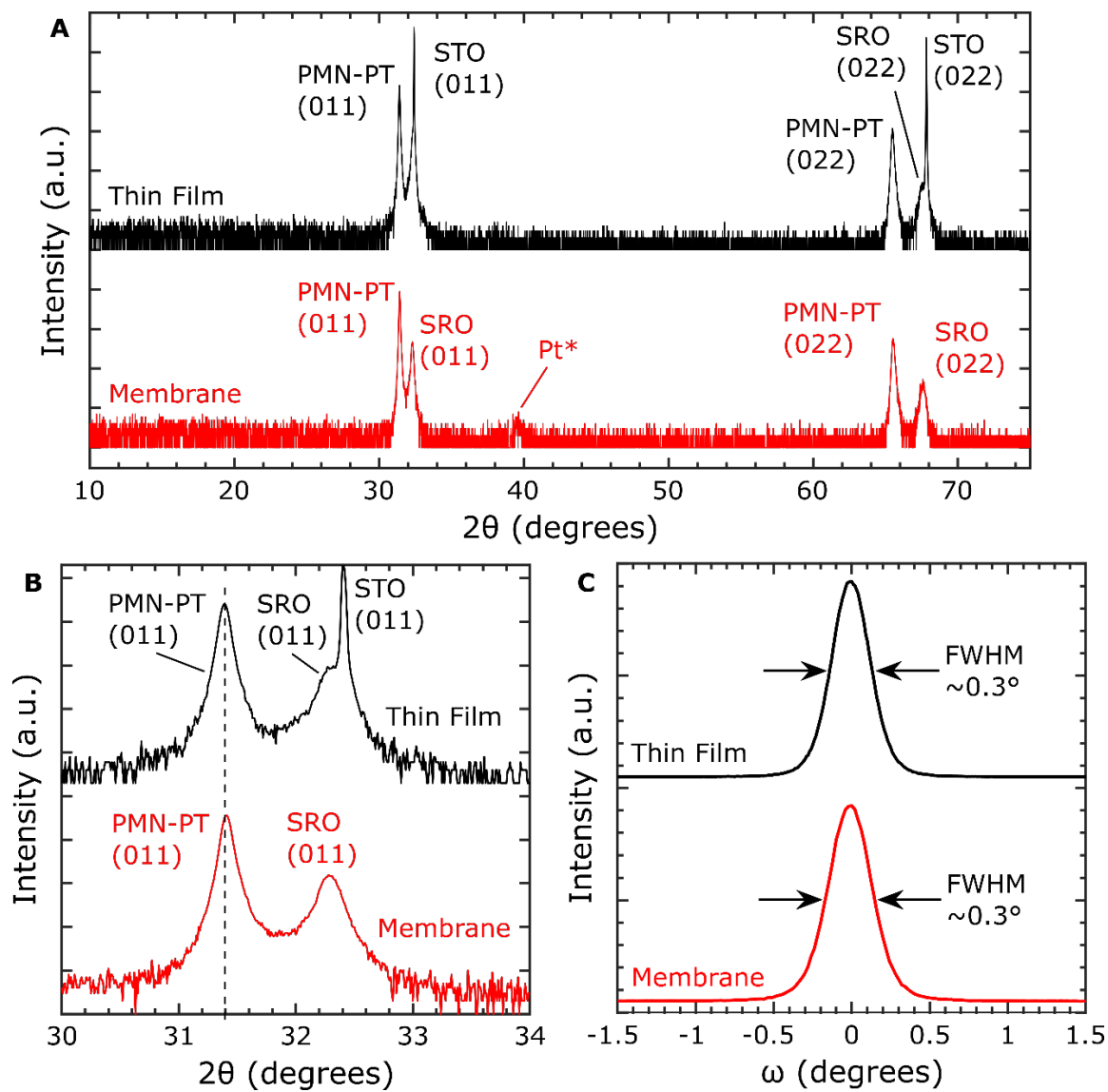

**Fig. S1. XRD structural characterization of PMN-PT thin film before and after release from substrate. (A)** θ-2θ scan showing that there are no secondary phases that appear in the PMN-PT thin film even after substrate removal. The SRO (011) peak is too close to the substrate to see in the film scan but appears after the substrate is released. There are no STO substrate peaks in the membrane scan due to the substrate having been completely removed. An additional Pt peak is visible in the PMN-PT membrane due to the deposition of Pt as the back electrode having been done before film release (see methods). **(B)** θ-2θ scan of the region around the (011) peaks. The SRO (011) peak is now visible in both the thin film and membrane scans. A vertical dashed line shows that the (011) PMN-PT peak does not shift much with position, indicating that upon release of the substrate the out-of-plane lattice parameter of the PMN-PT does not change significantly. **(C)** Phi scan of the PMN-PT (011) peaks in both before and after release look identical and have identical full width half maximums (FWHM), showing that the crystallinity of the film does not change after release from the substrate.

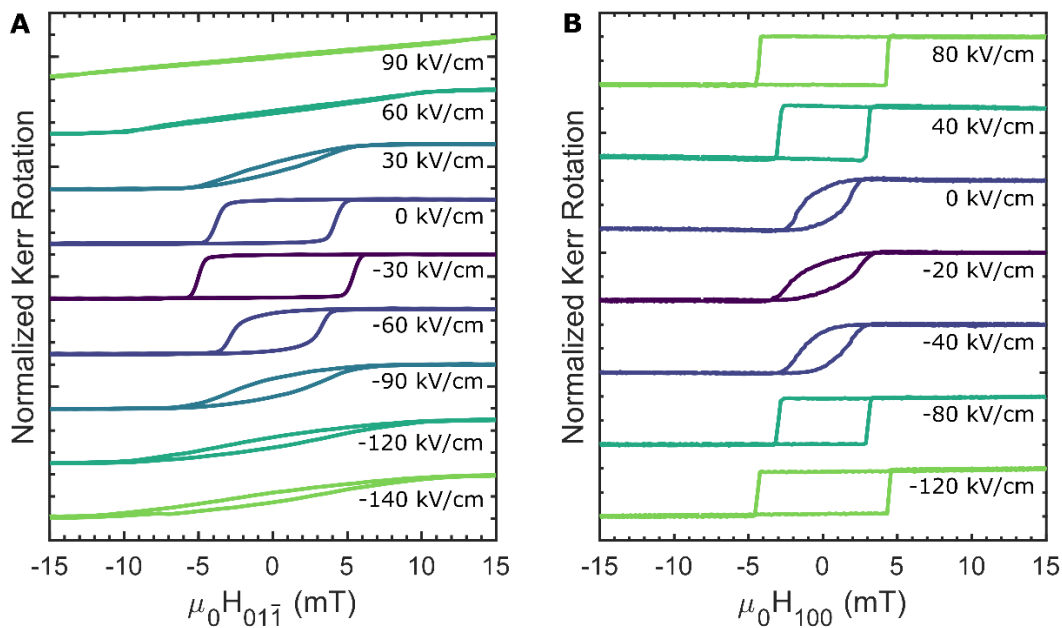

**Fig. S2. MOKE hysteresis loops with applied magnetic field along both [01$\bar{1}$] and [100].** MOKE hysteresis loops taken on the PMN-PT/Ni membrane with the magnetic field applied along [01$\bar{1}$] y-direction (A, same as in Figure 3A), as well as 90 degrees rotated in-plane along the [100] x-direction. At high fields, the square MOKE loops in (B) show that the EA anisotropy of the Ni is aligned along the x-direction due to compressive strains caused by polarization rotation towards the O$_{\text{UP}}$ direction (Figure 1). The slight tilting of the MOKE hysteresis at -20 kV/cm shows that the HA anisotropy is along the x-direction due to tensile strain along x.

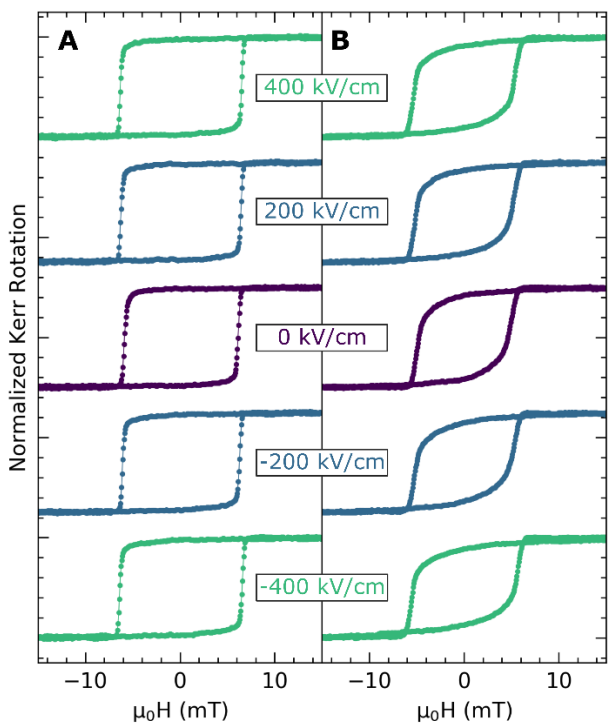

**Fig. S3. MOKE hysteresis loops with applied electric field in clamped Ni/PMN-PT thin films. (A)** Magnetic field aligned 30° from $x$ [100] direction. **(B)** Magnetic field aligned 30° from the $y$ [01$\bar{1}$] direction. No electric field induced anisotropy was present in clamped thin film samples, even up to 400 kV/cm applied bias (corresponding to 20V in our thin films). Samples used were Ni (35 nm) / PMN-PT (500 nm) / SrRuO$_3$ (100 nm) / SrTiO$_3$ and all films were sputtered as described in Methods. Ni top electrodes were patterned into 300 μm by 200 μm rectangles. Hysteresis loops of same electric field magnitude have the same color.

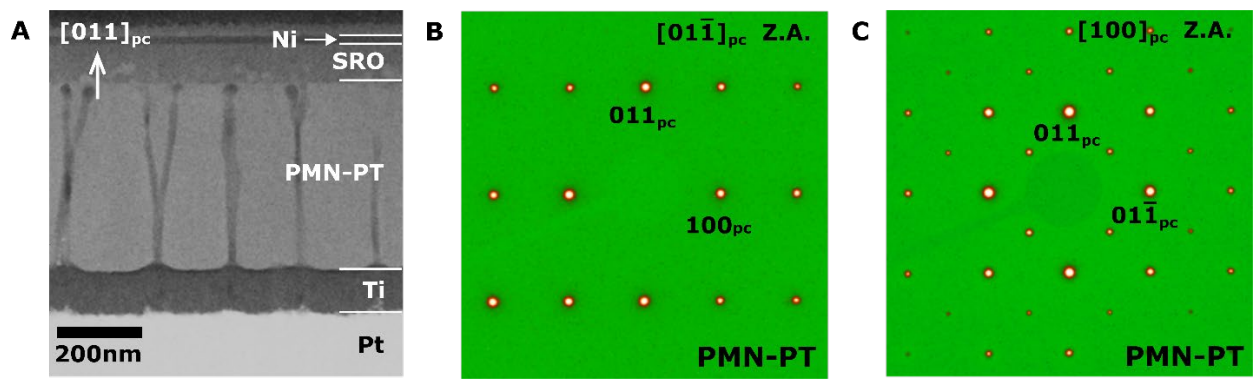

**Fig. S4. Cross-sectional structure image and selected area diffraction patterns of PMN-PT membrane. (A)** Cross section STEM image of the PMN-PT membrane heterostructure shows columnar structure. **(B-C)** Selected area diffractions of the PMN-PT membranes along the [01-1] and [100] pseudocubic zone axis show the membranes are single crystalline with a pseudocubic symmetry.

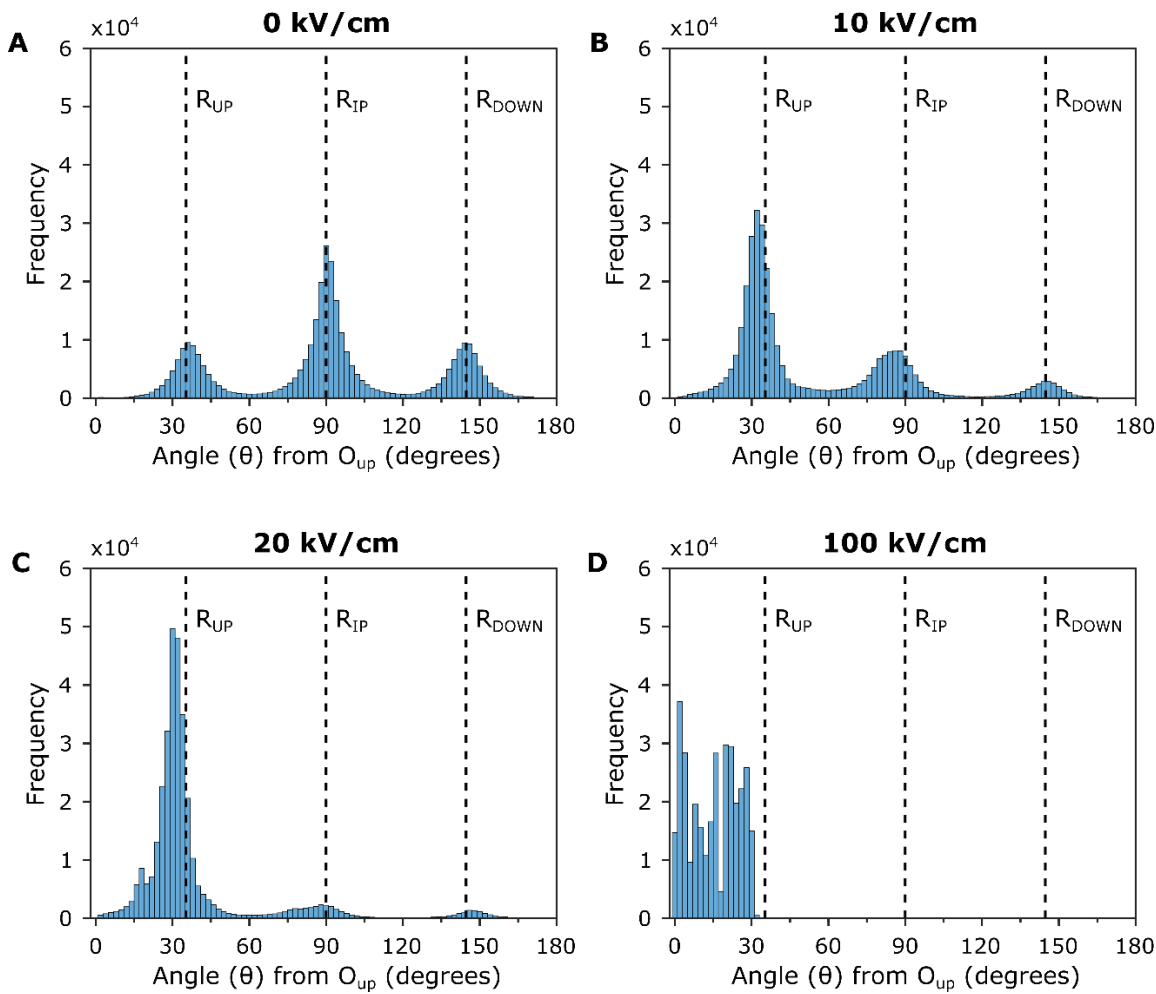

**Fig. S5. Phase-field simulation polarization angle histograms.** **(A)** At 0kV/cm, the simulation presents an even mixture of spontaneous polarizations along the R variants. **(B)** For a 10 kV/cm field applied along the z-direction [011], $R_{IP}$ and $R_{DOWN}$ domains switch to $R_{UP}$ domains. **(C)** By 20 kV/cm, almost all polarizations are near the $R_{UP}$ polarization direction, however, there is a shift towards a lower θ, indicating a monoclinic distortion towards the $O_{UP}$ direction. **(D)** In the high field region, all of the spontaneous polarizations are in the monoclinic region between $O_{UP}$ and $R_{UP}$, resulting in a large and negative differential in-plane strain. In each plot, the calculated angle for $R_{UP}$, $R_{IP}$, and $R_{DOWN}$ polarizations, assuming a pseudocubic unit cell, are shown by dashed lines.

**Supplementary Note S1 and Table S1: Calculated Polarization Group Strains**

The electrostriction tensor is used to calculate the strain arising from an electric polarization. These strains are relative to the paraelectric cubic phase. The strain tensor $\varepsilon_{ij}$ for a particular polarization vector $P_k$ and electrostriction tensor $Q_{ijkl}$ is

$$\varepsilon_{ij} = Q_{ijkl}P_kP_l \tag{1}$$

This equation may be written in the more compact Voigt matrix notation for cubic symmetry(*48, 49*)

$$\begin{pmatrix} \varepsilon_{11} \\ \varepsilon_{22} \\ \varepsilon_{33} \\ 2\varepsilon_{23} \\ 2\varepsilon_{31} \\ 2\varepsilon_{12} \end{pmatrix} = \begin{pmatrix} Q_{1111} & Q_{1122} & Q_{1122} & 0 & 0 & 0 \\ Q_{1122} & Q_{1111} & Q_{1122} & 0 & 0 & 0 \\ Q_{1122} & Q_{1122} & Q_{1111} & 0 & 0 & 0 \\ 0 & 0 & 0 & 4\,Q_{1212} & 0 & 0 \\ 0 & 0 & 0 & 0 & 4\,Q_{1212} & 0 \\ 0 & 0 & 0 & 0 & 0 & 4\,Q_{1212} \end{pmatrix} \begin{pmatrix} P_1^2 \\ P_2^2 \\ P_3^2 \\ P_2P_3 \\ P_1P_3 \\ P_1P_2 \end{pmatrix} \tag{2}$$

This can also be equivalently written with the commonly used matrix components $Q_{ij}$

$$\begin{pmatrix} \varepsilon_1 \\ \varepsilon_2 \\ \varepsilon_3 \\ \varepsilon_4 \\ \varepsilon_5 \\ \varepsilon_6 \end{pmatrix} = \begin{pmatrix} Q_{11} & Q_{12} & Q_{12} & 0 & 0 & 0 \\ Q_{12} & Q_{11} & Q_{12} & 0 & 0 & 0 \\ Q_{12} & Q_{12} & Q_{11} & 0 & 0 & 0 \\ 0 & 0 & 0 & Q_{44} & 0 & 0 \\ 0 & 0 & 0 & 0 & Q_{44} & 0 \\ 0 & 0 & 0 & 0 & 0 & Q_{44} \end{pmatrix} \begin{pmatrix} P_1^2 \\ P_2^2 \\ P_3^2 \\ P_2P_3 \\ P_1P_3 \\ P_1P_2 \end{pmatrix} \tag{3}$$

where $\varepsilon_1 = \varepsilon_{11}$, $\varepsilon_2 = \varepsilon_{22}$, $\varepsilon_3 = \varepsilon_{33}$, $\varepsilon_4 = \varepsilon_{23}$, $\varepsilon_5 = \varepsilon_{31}$, $\varepsilon_6 = \varepsilon_{12}$, $Q_{11} = Q_{1111}$, $Q_{12} = Q_{1122}$, and $Q_{44} = 2\,Q_{1212}$.

The 1, 2 and 3 indices of Eq. (2) refer respectively to the cubic crystal directions [100], [010], and [001]. Values for $Q_{ij}$ in Eq. (3) were taken from measurements reported in the literature on 28% rhombohedral PMN-PT crystals.(*12*) The strain tensor for a polarization group is found by equal weight averaging of the strain tensors calculated for each polarization vector present in the group.

The deformed unit cell for a particular polarization group is found from the principal strains of the strain tensor $\varepsilon_{ij}$ (constructed from the $\varepsilon_i$ and the symmetry condition $\varepsilon_{ij} = \varepsilon_{ji}$). The principal strains are three normal strains along directions given by the eigenvectors of $\varepsilon_{ij}$ with magnitudes given by the eigenvalues of $\varepsilon_{ij}$ . Each of the polarization group averages R$_{Up}$, R$_{IP}$, and O$_{Up}$ result in principal strain directions aligned with the $x$, $y$, and $z$ directions shown in Fig. 1A of the main text. The strain magnitudes along each direction, and the in-plane anisotropic strain $\varepsilon_{xx} - \varepsilon_{yy}$, are listed in Supplementary Table S1, both in terms of electrostriction matrix coefficients and numerically, with a 33 µC/cm$^2$ polarization magnitude. There is no shear strain since the strain tensor is by definition diagonal in the *xyz* coordinate system defined by the strain tensor eigenvectors.

**Supplementary Table S1 Electrostriction of polarization mixtures in (011)-oriented bulk PMN-PT. Strains were calculated analytically from $\varepsilon_{ij} = Q_{ijkl}P_kP_l$, and the 28% rhombohedral electrostriction tensor ($Q_{ijkl}$) values(*12*) were used to compute strains in ppm (shown in parentheses). Saturation polarization $P$ was taken to be 33 µC/cm². The frequently occurring term $Q_h = Q_{11} + 2Q_{12}$ is the hydrostatic electrostriction coefficient.(*50*)**

| Strain | $R_{IP}$ | $R_{Up}$ | $O_{Up}$ |
|---|---|---|---|
| $\varepsilon_{xx}$ | $P^2 \frac{1}{3}Q_h$ (254) | $P^2 \frac{1}{3}Q_h$ (254) | $P^2 Q_{12}$ (-2613) |
| $\varepsilon_{yy}$ | $P^2 \frac{1}{3}(Q_h + Q_{44})$ (1016) | $P^2 \frac{1}{3}(Q_h - Q_{44})$ (-508) | $P^2 \frac{1}{2}(Q_{11} + Q_{12} - Q_{44})$ (544) |
| $\varepsilon_{zz}$ | $P^2 \frac{1}{3}(Q_h - Q_{44})$ (-508) | $P^2 \frac{1}{3}(Q_h + Q_{44})$ (1016) | $P^2 \frac{1}{2}(Q_{11} + Q_{12} + Q_{44})$ (2831) |
| $\varepsilon_{xx} - \varepsilon_{yy}$ | $-P^2 \frac{1}{3}Q_{44}$ (-762) | $P^2 \frac{1}{3}Q_{44}$ (762) | $P^2 \frac{1}{2}(-Q_{11} + Q_{12} + Q_{44})$ (-3158) |

**Supplementary Note S2: Calculation of Ku and $\varepsilon_{xx} - \varepsilon_{yy}$**

In the case of a hard axis loop, the magnetic anisotropy energy density $K_U$ can be estimated using the experimental value of the full rotation field $H_{sat}$ to be

$$K_U = \frac{1}{2}\mu_0 M_{sat} H_{sat} \tag{4}$$

where $M_{sat}$ is the Ni saturation magnetization, assuming a coherent rotation model.(*51*) The strain induced anisotropy energy density is related to the Ni average anisotropic strain $\varepsilon_{xx} - \varepsilon_{yy}$ along principal strain axes by

$$K_U = \frac{3}{2}\lambda_s Y_{Ni}(\varepsilon_{xx} - \varepsilon_{yy}) \tag{5}$$

where $\lambda_s$ is the Ni magnetostriction coefficient (-33 ppm) and $Y_{Ni}$ is the Young's modulus of Ni (220 GPa).(*52*) Equations (4) and (5) allow the average anisotropic strain magnitude to be estimated from hard axis MOKE hysteresis loops. The sign of $\lambda_s$ for Ni dictates that $\varepsilon_{xx} - \varepsilon_{yy} > 0$ results in a magnetic easy axis along $[01\bar{1}]$, and $\varepsilon_{xx} - \varepsilon_{yy} < 0$ results in an easy axis along $[100]$.